\documentclass[final,5p,times,twocolumn]{elsarticle}
\usepackage[
    type={CC},
    modifier={by-nc-nd},
    version={4.0},
]{doclicense}

\usepackage{lineno,hyperref}
\usepackage{enumitem}
\usepackage{url}
\usepackage{glossaries}
\usepackage{graphicx}
\usepackage{subfig}
\usepackage{color}
\usepackage{todonotes}
\usepackage{ulem}
\usepackage{amssymb}

\graphicspath{{figures/}}

\journal{Astroparticle Physics}

\begin{document}

\binoppenalty=10000
\relpenalty=10000

\newcommand{\tmtexttt}[1]{{\ttfamily{#1}}}
\newcommand{\apj}{Astrophysical Journal}
\newcommand{\aap}{A\&A}
\newcommand{\pasp}{Publications of the ASP}

\definecolor{navyblue}{rgb}{0.0, 0.0, 0.5}
\newcommand{\thadd}[1]{\textcolor{navyblue}{[TH: #1]}}

\begin{frontmatter}

\title{Relevance of the fluorescence radiation in VHE gamma-ray observations with the Cherenkov technique}

\author[inst-gaeucm]{D. Morcuende\corref{cor1}}
       \ead{dmorcuen@ucm.es}
\author[inst-gaeucm]{J. Rosado\corref{cor1}}
       \ead{jaime\_ros@fis.ucm.es}
\author[inst-gaeucm]{J.L. Contreras}
\author[inst-gaeucm]{F. Arqueros}

\address[inst-gaeucm]{High Energy Physics Group GAE and UPARCOS. Facultad de Ciencias F\'{i}sicas. Universidad Complutense de Madrid. E-28040 Madrid, Spain.}

\cortext[cor1]{Corresponding author}

\begin{abstract}
Atmospheric fluorescence is usually neglected in the reconstruction of the signals registered by Cherenkov telescopes, both IACTs and wide-angle detector arrays. In this paper we quantify the fluorescence contribution to the total light recorded in typical observational configurations. To this end we have implemented the production and tracking of fluorescence light in the CORSIKA code. Both the Cherenkov and fluorescence light distributions on ground (2200~m a.s.l.) have been simulated for $\gamma$-ray showers in a wide energy range ($10^{-1} - 10^3$ TeV). The relative fluorescence contribution has been evaluated as a function of the shower energy and zenith angle. Our results indicate that at distances from the impact point smaller than 200~m the fluorescence contamination is negligible in both types of telescopes. However, at about 1000~m, the contamination in IACTs is around 5\%. At these core distances, the signals recorded by wide-angle detectors contain around 45\% of atmospheric fluorescence in the PeV region.
\end{abstract}


\end{frontmatter}

\section{Introduction}
\label{sec:introduction}
The charged particles of extensive air showers (EASs) generate light due to two physical processes; namely, the Cherenkov effect and the fluorescence emission. The detection of the atmospheric Cherenkov radiation is the basis of several observational techniques, the imaging atmospheric Cherenkov telescopes (IACTs) being the most powerful one for very high-energy (VHE) $\gamma$-ray astronomy. On the other hand, the observation of atmospheric fluorescence with optical telescopes has been used since many years to measure the longitudinal development of EASs in the ultra-high-energy (UHE) range ($>10^{17}$~eV).
\par
The Cherenkov and fluorescence photons emitted from any point in an EAS are indistinguishable at the telescope level because they arrive simultaneously and within the same wavelength range of around $300-450$~nm. However the angular distributions of both components are quite different. The Cherenkov radiation is peaked at a small angle of $\sim 1^{\circ}$ around the shower axis, forming on ground the so-called {\it Cherenkov light pool}. On the contrary, the fluorescence emission is isotropic. As a consequence, the signal of a telescope pointing towards the shower axis is dominated by the Cherenkov component, while the signal of a telescope recording the shower transversely is dominated by the fluorescence one. Another difference between both processes is that fluorescence emission is less efficient than Cherenkov production. For instance, a 1~GeV electron in 1~m of atmosphere near ground generates about 30 Cherenkov photons but only about 4 fluorescence photons.
\par
In the IACT technique, one or more telescopes with a field of view (FoV) of some degrees point at a given $\gamma$-ray source from different positions. Fast Cherenkov pulses are recorded to image the EASs stereoscopically (see, e.g., \cite{Hinton2009,Magic2010}). We single out the Cherenkov Telescope Array (CTA), which is the next-generation IACT observatory currently entering into its construction phase and designed to cover the wide energy range from 20~GeV to 300~TeV \cite{TheCTAConsortium2013a,cta-performance}. The atmospheric Cherenkov light can also be registered by arrays of wide-angle Cherenkov detectors (WACD). Each unit usually consists of one or more photomultipliers that collect the light through a Winston cone of $\sim 1$~sr acceptance. These facilities are used as non-imaging $\gamma$-ray telescopes with an energy threshold of a few tens of TeV and large sky coverage as well as for cosmic-ray studies over 0.1~PeV \cite{AIROBICC, Tluczykont2014, Gress2013Tunka-HiSCOREPhysics}. Apart from the direct Cherenkov light, scattered Cherenkov photons as well as fluorescence photons contribute to the signals recorded by both imaging and non-imaging $\gamma$-ray telescopes. The contribution of scattered light has been proved to be marginal \cite{Bernlohr2000}. The fluorescence light is also assumed to be negligible, but no systematic study has been made to quantify it yet.
\par
Fluorescence telescopes are non-steerable devices with a wide FoV that register transversely the track of the shower in the atmosphere within a typical time window of microseconds. Although the signal is dominated by the fluorescence component, a non-negligible amount of both direct and scattered Cherenkov light gets into the telescope and it has to be included as a relevant ingredient in the reconstruction algorithm \cite{Unger2008ReconstructionMeasurements}. The fluorescence technique has been successfully employed by the HiRes project \cite{Abu-Zayyad2000}, the Pierre Auger Observatory \cite{Auger2010} and the Telescope Array project \cite{Tokuno2012}. Besides, the LHAASO project has designed flexible and mobile telescopes that can be operated in both Cherenkov and fluorescence modes \cite{Zhang2011-LHAASO}.
\par
A detailed Monte Carlo (MC) simulation of the development of EASs is a key tool in this field. Notably CORSIKA \cite{corsika} has been extensively used since the late 90's for this purpose. Cherenkov telescopes need detailed simulations of the Cherenkov radiation generated by the shower. To this end, the Cherenkov production was implemented in CORSIKA  for the simulation of the HEGRA AIROBICC detector \cite{Martinez1995} and later on upgraded by K. Bernl\"ohr \cite{Bernlohr2008}. The simulation of the fluorescence telescope signals is customarily based on the generation of both Cherenkov and fluorescence photons from a one-dimension shower profile obtained by either an analytical EAS model or a MC code like CORSIKA (see, e.g., \cite{PradoJr.2005}).
\par
In a previous work \cite{ICRC17}, we made a preliminary implementation of the fluorescence emission in CORSIKA, following a procedure similar to that used by V. de Souza et al. \cite{DeSouza2004a}. We showed that the fluorescence contribution to the signals of Cherenkov telescopes might be non-negligible in some particular cases. In this paper we present an updated implementation of the fluorescence emission and systematic MC results to quantify this fluorescence contamination in ground-based Cherenkov telescopes.
\par
The paper is structured as follows. In section \ref{sec:fluorescence}, the atmospheric fluorescence in EASs and the used parameterization are described. In section \ref{sec:CORSIKA}, we detail the implementation of the fluorescence emission in CORSIKA. The MC production and the tools used to extract the relevant information are described in section \ref{sec:MC_simul}. In section \ref{sec:results}, the results on fluorescence contamination in both IACTs and WACDs are shown for typical observational conditions. Also a cross-check of our MC predictions with a numerical algorithm is presented. In section \ref{sec:future_work}, foreseen improvements of our fluorescence implementation in CORSIKA as well as other potential applications in $\gamma$-ray astronomy and cosmic rays physics are discussed. Finally, in section \ref{sec:conclusions}, we present the conclusions of this work.

\section{Atmospheric fluorescence in extensive air showers}
\label{sec:fluorescence}
Atmospheric fluorescence\footnote{This radiation should be named scintillation, although for historical reasons the cosmic-ray community refers to it as fluorescence.} is produced by the de-excitation of molecules previously excited or ionized by charged particles of an EAS \cite{Arqueros2009}. In the near UV, molecular nitrogen is mainly responsible for this light emission, which is dominated by the so-called Second Positive system (2P) of N$_2$ (C$^3\Pi_u \rightarrow $ B$^3\Pi_g$) and the First Negative system (1N) of N$^+_2$ (B$^2\Sigma^+_u \rightarrow$ X$^2\Sigma^+_g$). Each one consists in a number of molecular bands according to the vibrational quantum numbers $v', v$ of the upper and lower electronic states of the transition, respectively. A small contribution from the Gaydon-Herman system of N$_2$ is also present \cite{AIRFLY2007}. Shower electrons and positrons are mostly responsible for the fluorescence emission, while other charged particles, if produced (e.g., in hadronic showers), contribute to a minor extent. 
\par
Although the radiative lifetimes of involved molecular states are of some tens of nanoseconds, the emission in the lower atmosphere is shortened down to the ns level (e.g., $\approx$ 1~ns at 7000~m a.s.l. and $\approx$ 2~ns at 12000~m a.s.l.). This effect is due to collisional quenching, that is, excited molecules can also relax without emitting light through collisions with other molecules, oxygen and water vapor being the dominant quenchers. The collisional cross section depends on the kinetic energy of the encounters. As a consequence, the light intensity generated by an electron moving in the air depends on the atmospheric conditions (i.e., pressure, temperature and humidity). 
\par
The fluorescence yield, defined as the number of photons emitted per unit of deposited energy, is a key parameter to quantify the fluorescence emission. It has been shown that this parameter is basically independent of the energy and type of
ionizing particle \cite{Rosado2014}, that is, the fluorescence emission is proportional to the energy deposit at given atmospheric conditions. For a molecular band at a wavelength $\lambda$, the fluorescence yield at pressure $P$ can be expressed as
\begin{equation}
\label{FY1}
Y_{\lambda}=\frac{Y^0_{\lambda}}{1+P/P'_{\lambda}}\,,
\end{equation}
where $Y^0_{\lambda}$ is the fluorescence yield in the absence of collisional quenching (i.e., when $P \rightarrow 0 $). The parameter $P'_{\lambda}$ is defined as the pressure for which the probability of radiative de-excitation equals that of collisional quenching and depends on temperature $T$ and humidity $h$.
\par
Alternatively, the fluorescence yield for any molecular band can be expressed relative to the absolute yield measured for a reference band in dry air at a given pressure $P_0$ and temperature $T_0$ as
\begin{equation}
\label{parametrization}
Y_{\lambda}(P,T,h)=Y_{337}(P_0,T_0)\,I_{\lambda}(P_0,T_0)
\frac{1+P_0/P'_{337}(T_0)}{1+P/P'_{\lambda}(T,h)}\,,
\end{equation}
where the dependencies on the atmospheric parameters are now put explicitly. Here, we use as reference band the one located at 337~nm wavelength, which is the most intense band of the air fluorescence spectrum and corresponds to the 2P ($v'=0 \rightarrow v=0$) transition. The world average $Y_{337}=7.04\pm 0.24$~MeV$^{-1}$ at 800~hPa and 293~K in dry air reported in \cite{Rosado2014} has been used in this work. The intensities $I_{\lambda}(P_0,T_0)$ of the other bands in the interval $284-429$~nm relative to this reference one at the same air conditions were taken from \cite{AIRFLY2007}. The $P'_{\lambda}(T,h)$ parameters, including the one of the reference band $P'_{337}(T_0)$ for dry air, were those given in \cite{AIRFLY2007} and \cite{AIRFLY2008}. The wavelength interval for these calculations was extended to $280-700$~nm to include some other weak bands of the 2P and 1N systems, for which we used the corresponding branching ratios and assumed $P'\propto T^{1/2}$ \cite{Arqueros2009}. The effect of the selection of air-fluorescence parameters in the reconstruction of EASs has been studied in \cite{Vazquez2013,Vazquez2013a}.

\section{Implementation of fluorescence in CORSIKA}
\label{sec:CORSIKA}

Firstly, in subsection \ref{subsec:work-flow}, the details of particle transportation relevant for fluorescence and Cherenkov emission as well as the main features of the Cherenkov implementation in CORSIKA are summarized. Later on, our implementation of the fluorescence production is described in subsection \ref{subsec:implement_fluor}.

\subsection{Particle transportation and Cherenkov radiation}
\label{subsec:work-flow}
CORSIKA handles the transport of all particles generated from the very first interaction of the incoming particle down to the observation level. They are tracked while their energy is above a certain threshold, which is set by the user for each particle type. In addition, particles going upwards are discarded from being tracked\footnote{The transportation of upward-going particles may be activated in CORSIKA, but this option is not included in the default settings, as it is irrelevant for the determination of the Cherenkov light at ground level.}. When a particle is propagated between two interaction points, its spatial and time coordinates are updated. For charged particles, the ionization energy loss, the multiple scattering and the bending of the trajectory in the Earth's magnetic field are also evaluated. Once the type of inelastic interaction or decay that the particle undergoes is determined, the outcoming particles are tracked, provided that they survive the above-mentioned energy and angular cuts.
\par
The Cherenkov radiation production within a predefined wavelength interval is evaluated each time a charged particle fulfilling the Cherenkov condition undergoes a transportation step. The step is subdivided into smaller sub-steps of equal length and a bunch of photons, treated as a whole, is generated at the midpoint of each sub-step. The number of sub-steps of a given transportation step is chosen such that the number of photons in any bunch\footnote{In general, this is a fractional number.} is always less than a certain {\it bunchsize} set by the user. For each bunch, the number of photons and their emission angle relative to the particle direction are determined by the refractive index at the corresponding height and the velocity of the particle, which is calculated taking into account the continuous energy loss along its trajectory. The azimuthal emission angle of the bunch around the particle direction is taken at random. Finally the bunch is transported to ground level and its information (i.e., number of photons, spatial and time coordinates at observation level, arrival direction and height of emission) is written to a file as long as it reaches the detector area defined by the user. This procedure reduces significantly the computation and storage needs, while includes statistical fluctuations in the spatial distribution of bunches on ground.
\par
Optionally, CORSIKA can assign a specific wavelength to each bunch to evaluate both the refraction and the absorption of the atmosphere as well as the photon detection efficiency of the telescopes. However, this option needs substantial computational resources and is not activated in the default settings of CORSIKA.

\subsection{Air-fluorescence emission}
\label{subsec:implement_fluor}
The fluorescence emission was implemented in the CORSIKA v7.6300 code. We followed the same scheme of photon bunches used in the simulation of the Cherenkov light production described above. Every time a charged particle deposits energy in the atmosphere, bunches of fluorescence photons are generated. The amount of fluorescence light is evaluated within a wavelength interval set by the user. In the workflow of CORSIKA the energy deposition takes place under two different circumstances that are described next, together with the procedure to build-up the bunches. 
\par
In the first place, when a particle is transported between two interaction points, it deposits an amount of energy equal to the continuous ionization energy loss. Most of the charged particles of an EAS are electrons and positrons having transportation steps almost always shorter than 20~m, for which the atmospheric characteristics can be assumed to be constant. For very penetrating particles, e.g., muons, with much larger transportation steps, these are split into segments of $\leq$ 20~m length\footnote{This maximum step size has been chosen somewhat arbitrarily, inspired in the maximum typical step lengths for electrons. Probably it could be chosen larger without compromising the accuracy of the simulation, but this fine step splitting does not increase significantly the computing time because it is applied in a very small fraction of cases.} and the total energy deposit is distributed proportionally to the thicknesses of the newly defined steps.
\par
The number of fluorescence photons in each step is randomly calculated from a Poisson distribution with mean equal to the photon yield calculated as described in section \ref{sec:fluorescence}, taking the atmospheric parameters at the step midpoint\footnote{The standard atmospheric parameterization in CORSIKA only has information on the air density and the mass overburden as a function of height. Therefore, pressure and temperature were calculated under the assumption of an ideal gas under a gravitational acceleration of 9.81~m/s$^2$. Humidity effects were neglected.}. Then, photons are equally distributed in bunches, each one generated in a sub-step in the same way as for the Cherenkov light production. However the maximum fluorescence bunchsize is defined independently to allow for a different granularity of these two processes. Each bunch is emitted at a random direction, but restricted to downward angles to save computing time.
\par
The other circumstance when energy is deposited is that a particle does not survive the angular or energy cuts. Note that, as opposed to the Cherenkov light production, this has to be considered for the fluorescence production, because this process has no sizable energy threshold and is isotropic. The energy deposit is assumed to occur at the point where CORSIKA stops tracking the particle, and it depends on the type of particle. Photons, electrons and other stable particles deposit their kinetic energy. For positrons, the annotated energy deposit includes additionally the energy corresponding to two electron masses released as annihilation photons, which are not tracked. For muons and other unstable particles, the energy deposit is the total particle energy minus the mean energy fraction carried by neutrinos, which are supposed to deposit no energy in the atmosphere.
\par
Each time CORSIKA applies the angular or energy cuts to a particle, bunches of fluorescence photons are emitted at random directions from the point where the energy is deposited. The distribution of photons in bunches is done taking into account the fluorescence bunchsize, as explained before.
\par
Some remaining small contributions due to a variety of hadronic interactions producing subthreshold secondary particles are not included yet, but they were checked to be less than 1\% of the total energy deposited by electromagnetic showers and less than 3\% for EASs induced by TeV protons.
\par
In the same way as for the Cherenkov radiation, only bunches reaching the predefined detector area are stored. The information of both Cherenkov and fluorescence bunches are written to the same standard output file. We made use of the sign bit of the variable that stores the number of photons to differentiate between the two types of bunches.

\section{Monte Carlo simulations}
\label{sec:MC_simul}
The implementation presented in the previous section has been applied to simulate the fluorescence light at conditions relevant for Cherenkov telescopes, with the aim of studying the importance of the fluorescence contamination. Next, we describe the used CORSIKA setup and input parameters as well as the tools that we developed to analyze the simulation data.
\par
In this work we used the high- and low-energy hadronic interaction models of \texttt{QGSJET-01C} and \texttt{GHEISHA 2002d}, respectively. The \texttt{CERENKOV} option, needed for the simulation of Cherenkov radiation, was used in combination with the EGS4 package that allows the full Monte Carlo treatment of the electromagnetic component. The energy thresholds were set to 0.3~GeV for hadrons, 0.1~GeV for muons and 0.02~GeV for the electromagnetic component, which are typical values for the simulation of Cherenkov telescopes
and sufficient for the purposes of this work.
\par
The detector was defined as a $5 \times 5$ km$^2$ square. As a representative case, we set the observation altitude at 2200~m a.s.l., which is the one of the Roque de los Muchachos Observatory, island of La Palma. The geomagnetic field corresponding to this location was introduced. The US Standard atmospheric profile was used to simulate the fluorescence and Cherenkov emission, but the atmospheric absorption was not included. According to CORSIKA coordinate system, the $x$ axis points to the magnetic North and the $y$ axis towards the West direction.
\par
The maximum bunchsize for Cherenkov and fluorescence light emission was set to 10 and 5, respectively. In this way we compensated the more efficient production of Cherenkov photons with respect to that of fluorescence and the isotropic emission of the latter. The wavelength range of the simulated photons
was $300-450$~nm for both light components.
\par
\begin{figure}[!htb]
\begin{center}
\centering\includegraphics[width=0.7\linewidth]{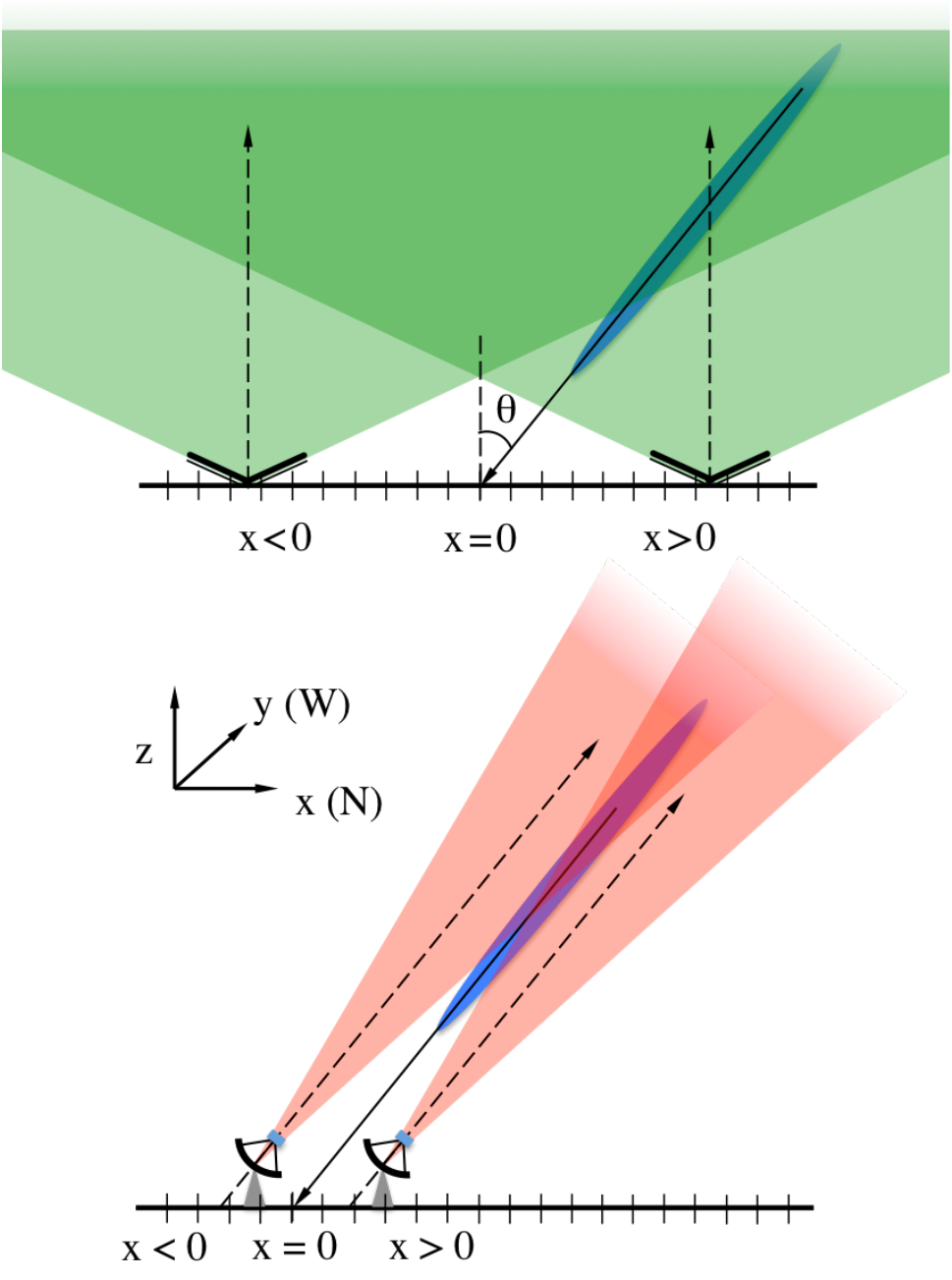}
\caption{Geometry and coordinate system adopted in the simulations. Inclined showers are assumed to go from North to South. Light density on ground is binned along the $x$ axis, assuming that the shower core impacts at the origin. The FoVs of two wide-angle Cherenkov detectors (WACDs) and two imaging air-Cherenkov telescopes (IACTs) are illustrated in the top and bottom plots, respectively.}
\label{fig:geometry}
\end{center}
\end{figure}

The MC production carried out for this work consists in $\gamma$-ray showers within a large energy range (100~GeV to 1~PeV) with zenith angles $\theta$ between $0^\circ$ and $60^\circ$. To get a good statistics of the photon density distributions on ground, the number of events was scaled with the shower energy (e.g., 1000 showers of 10~TeV and 100 showers of 100~TeV for each $\theta$ value). Simulated showers were forced to impact at the center of the detector area so that arrival coordinates of bunches are given relative to the shower core at observation level. Inclined showers were assumed to go always from North to South, as depicted in figure \ref{fig:geometry}.

\begin{figure*}[!htb]
\begin{center}
\centering\includegraphics[width=0.8\linewidth]{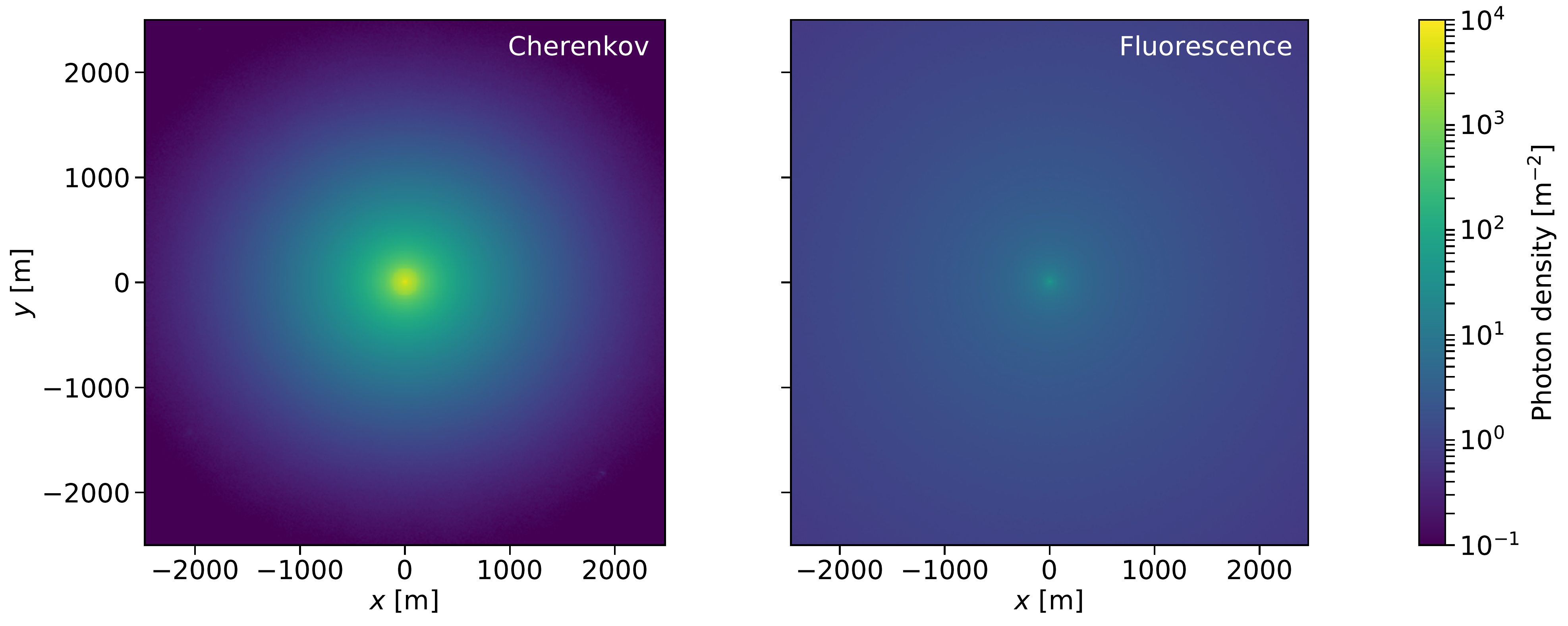}
\caption{Averaged spatial distributions of Cherenkov (left) and fluorescence (right) light on ground from 10 TeV $\gamma$-ray vertical showers.}
\label{fig:10TeV_vertical_2D}
\end{center}
\end{figure*}

\begin{figure}[!htb]
\begin{center}
\centering\includegraphics[width=0.9\linewidth]{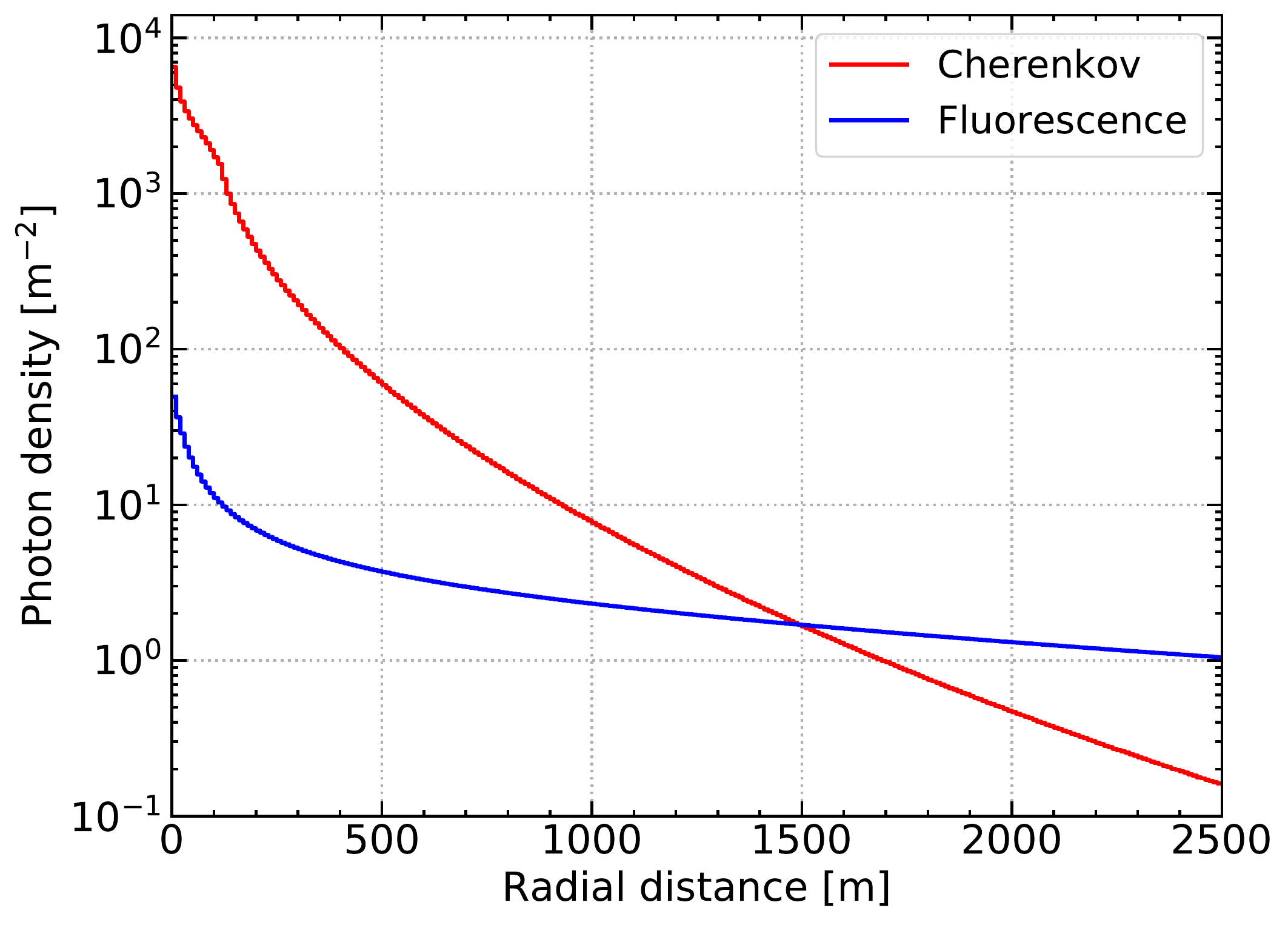}
\caption{Radial distributions of Cherenkov (red) and fluorescence (blue) light obtained after averaging the 2D distributions of figure \ref{fig:10TeV_vertical_2D} over azimuthal angle.}
\label{fig:10TeV_vertical_radial}
\end{center}
\end{figure}

Dedicated Python-based scripts were developed to read the output files containing the bunch information and calculate the spatial distributions of Cherenkov and fluorescence photons on ground. For this purpose, the detector area of $5 \times 5$ km$^2$ was discretized into a dense grid and the number of photons in each grid element was counted, i.e., summing over all the bunches impinging on it. Cuts on the arrival angle of the bunches were implemented in such a way that the pointing direction and FoV of a Cherenkov telescope can be simulated (see figure \ref{fig:geometry}). No cut in the arrival time of bunches was applied for this work, although it can be easily implemented for further studies of the time profile of the recorded signals.

\section{Results}
\label{sec:results}
In subsection \ref{subsec:general_features}, the spatial distribution of both Cherenkov and fluorescence light components on ground are compared with the assistance of some illustrative examples. Afterwards, we present the results of a systematic study on the fluorescence contamination in Cherenkov telescopes as a function of the primary $\gamma$-ray energy (subsection \ref{subsec:energy}) and zenith angle (subsection \ref{subsec:zenith_angle}). For the IACT technique, the impact of the observation off-set angle is analyzed in subsection \ref{subsec:FoV}. Finally, in subsection \ref{subsec:algorithm}, our implementation of the fluorescence emission is cross-checked against the predictions from a numerical algorithm.    

\subsection{General features of the light distribution on ground}
\label{subsec:general_features}
As an illustrative case, the spatial distribution of both Cherenkov and fluorescence photon densities averaged over 1000 vertical 10~TeV $\gamma$-ray showers is shown in figure \ref{fig:10TeV_vertical_2D}. For this example, no angular cuts have been applied. Both distributions have azimuthal symmetry around the shower core position, as the effect from the geomagnetic field is very small. Thus, we averaged the Cherenkov and fluorescence photon densities over azimuth angle to obtain the corresponding radial profiles (figure \ref{fig:10TeV_vertical_radial}). Cherenkov light concentrates on the pool region, which extends up to a radial distance of $\sim 120$~m in this case\footnote{The extension of the Cherenkov light pool on ground depends on the shower inclination and observation altitude, but it is almost independent of the primary $\gamma$-ray energy.}, while the photon density sharply drops outside this region. On the other hand, the fluorescence photon density decreases much less steeply with increasing core distance due to the isotropic nature of the fluorescence emission. In fact, the fluorescence photon density becomes even greater than the Cherenkov one at very large core distance.

\begin{figure}[ht]
\begin{center}
\centering\includegraphics[width=0.9\linewidth]{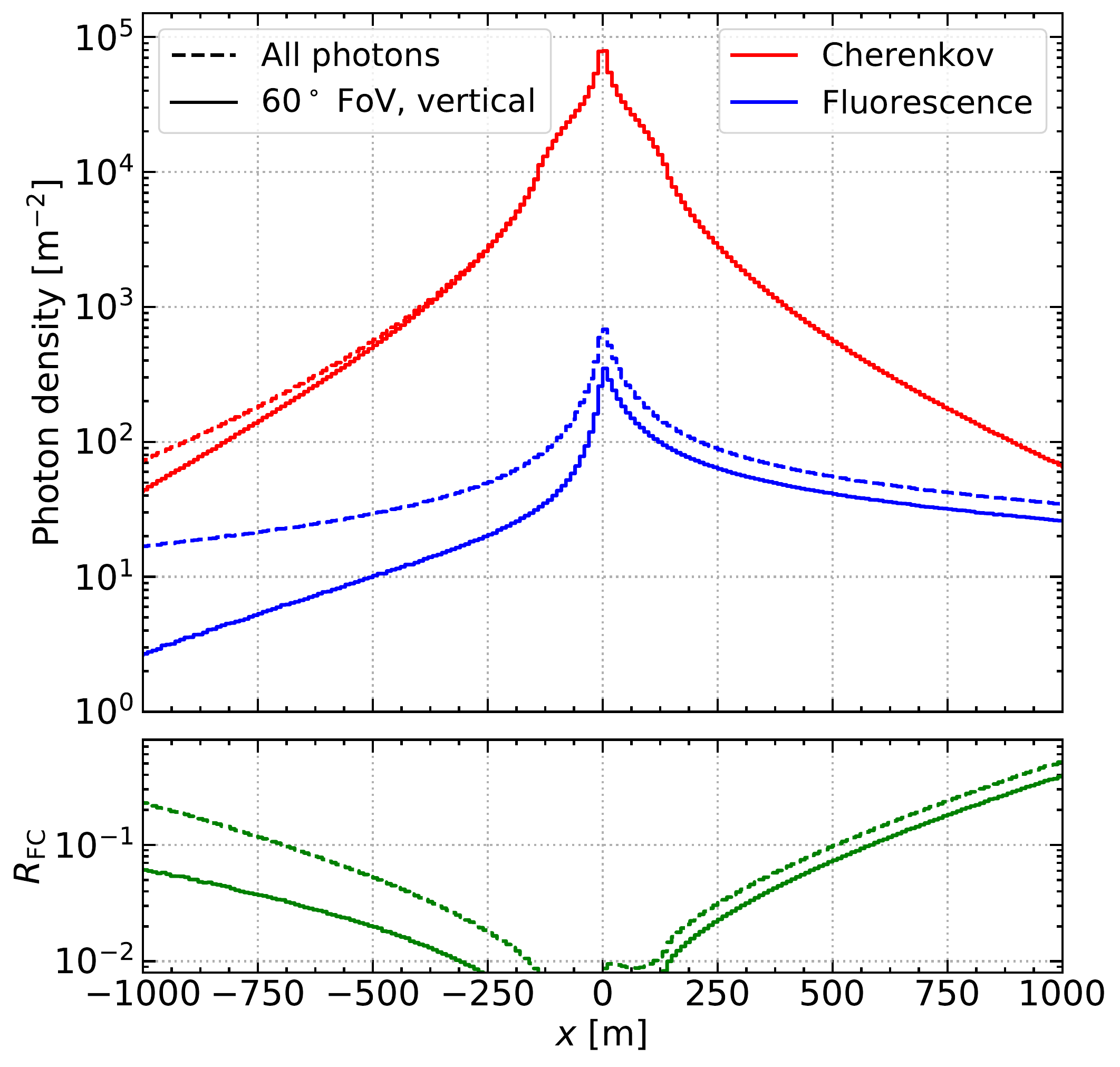}
\caption{Upper panel: Averaged lateral profiles of Cherenkov and fluorescence photons with no angular cut (dashed lines) and with zenith angles lower than $30^\circ$ (solid lines) from 100~TeV $\gamma$-ray showers with an inclination angle of $20^\circ$. Lower panel: Fluorescence contamination versus distance to the impact point.}
\label{fig:profiles_wide_angle}
\end{center}
\end{figure}

In a general case of inclined showers, the photon density distributions on ground do not have azimuthal symmetry. For simplicity, we calculated the photon density profiles only along the $x$ axis. Therefore, we restricted the detection area to a strip of 10~m width along this axis and defined bins of $10\times 10$~m$^2$ within this strip (see figure \ref{fig:geometry}). Results at arbitrary $(x,\,y)$ positions can be estimated from geometrical considerations, as shown in section \ref{subsec:zenith_angle}.
\par
\begin{figure}
\begin{center}
\centering\includegraphics[width=0.9\linewidth]{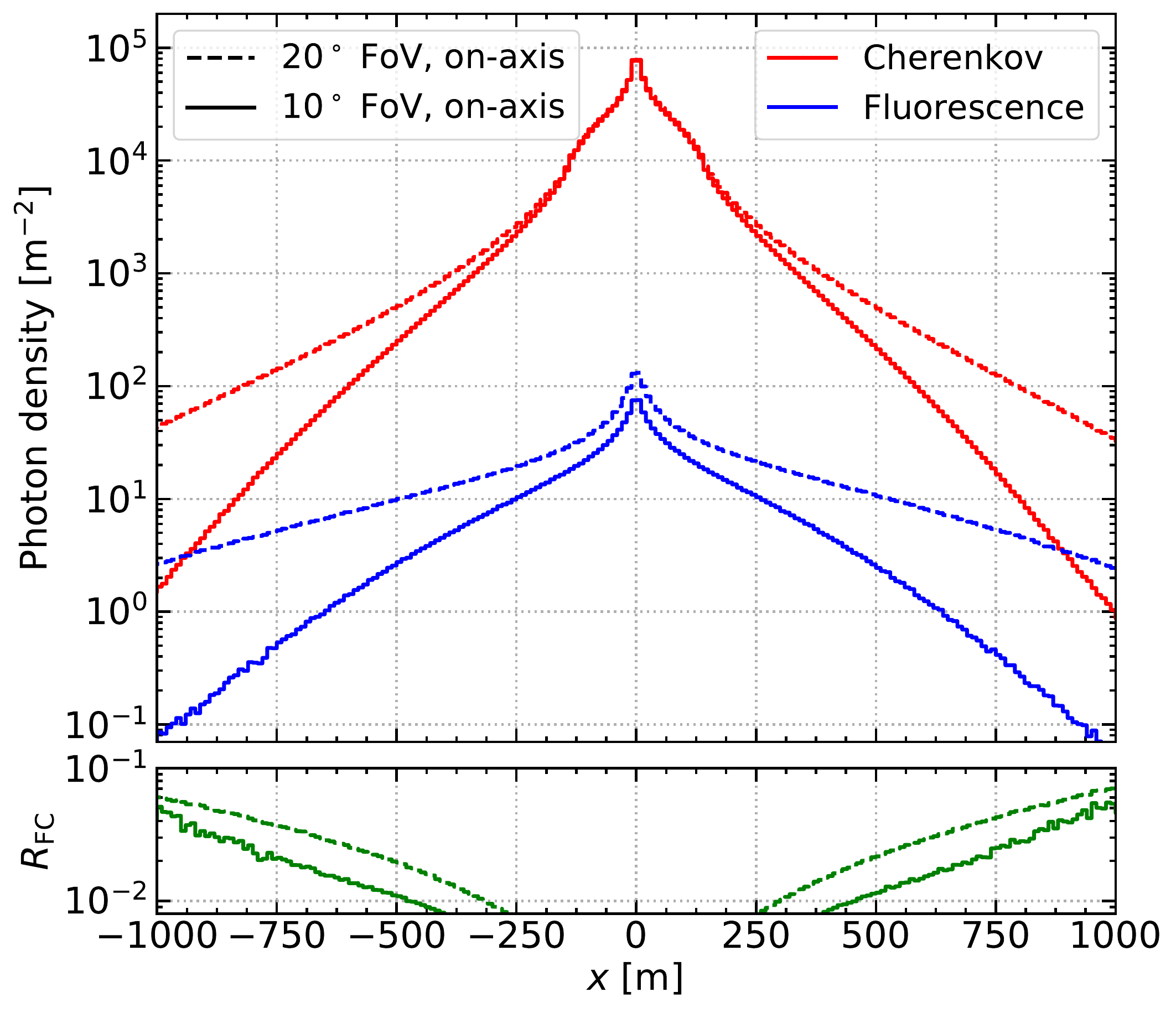}
\caption{The same as figure \ref{fig:profiles_wide_angle}, but for IACTs with two hypothesis of FoV: $10^\circ$ and $20^\circ$ in diameter, and on-axis observation.}
\label{fig:profiles_IACT}
\end{center}
\end{figure}

Examples of the photon density profiles for 100~TeV showers with $\theta=20^\circ$ averaged over 100 MC events are shown in the upper panels of figures \ref{fig:profiles_wide_angle} and \ref{fig:profiles_IACT}. Two distinct observational techniques are contemplated. Figure \ref{fig:profiles_wide_angle} illustrates the case of WACDs pointing to the zenith, like those of \cite{AIROBICC, Tluczykont2014, Gress2013Tunka-HiSCOREPhysics}. Here, the dashed lines represent the light profiles without any angular cut (i.e., assuming a wide open FoV of $2\,\pi$~sr), while the solid lines correspond to the profiles obtained by imposing the realistic condition that photons arrive with a zenith angle lower than $30^\circ$ (i.e., within a FoV of 0.85~sr around the vertical direction). Without angular cuts, the fluorescence photon density is smaller for negative $x$ values due to
the geometric attenuation factor $\frac{1}{d^2}$ of the isotropic fluorescence emission, where $d$ is the distance that separates the telescope and a given emission point in the shower.
On the contrary, the distribution of Cherenkov light has an almost perfect mirror symmetry due to its directional emission. When the cut on zenith angle is applied, the Cherenkov photon density remains basically unchanged for $\lvert x\rvert<1000$~m, because most of the Cherenkov light is emitted with small angles with respect to the shower axis and thus almost all Cherenkov photons reaching the WACD come from the part of the shower that is within the FoV. The contribution of the lower part of the shower outside the detector FoV can however be important for the fluorescence light component due to its isotropic nature. The fraction of fluorescence photons passing the angular cut is around 70\% for positive $x$ values, while it is significantly lower along the negative $x$ axis, since the portion of the shower missed by the WACD is larger when observing from that side, as illustrated in the top panel of figure~ \ref{fig:geometry}. For a typical detector area of the order of 0.1~m$^2$, the trigger level is around $10^3$ photons per square meter \cite{Gress2013Tunka-HiSCOREPhysics}. According to the photon densities shown in the figure, this translates into a maximum acceptable $x$ value of a few hundreds of meters for 100~TeV showers, although core distances up to 1000~m can be of interest in the PeV energy range.
\par
The scenario shown in figure \ref{fig:profiles_IACT} is that of IACTs with a FoV of $10^\circ$ (solid lines) or $20^\circ$ (dashed lines) in diameter assuming on-axis observation (see subsection \ref{subsec:FoV} for a discussion about off-axis effects). Narrower FoVs, e.g., $5^\circ$ in diameter, are only used in the low-energy range ($<1$~TeV) and with small impact parameters for which the fluorescence contamination is negligible. The photon density was calculated for a horizontal surface element regardless of the inclination of the telescope. Core distances greater than 1000~m were not considered in the IACT case either, because the light intensity collected by a telescope positioned at such a distance would be extremely low and it would only image the uppermost portion of the shower. Notice that the lateral distributions exhibit a strong mirror symmetry, because the part of the shower inside the FoV is almost the same for both sides, i.e., $x<0$ and $x>0$ (see bottom panel of figure \ref{fig:geometry}). The narrower the FoV, the smaller the portion of the shower development within it and the lower the fraction of both Cherenkov and fluorescence photons detected at a given core distance. However, the reduction in photon density is more important for the fluorescence component, since the missed portion of the shower is the lowest one.

In this work, we define the fluorescence contamination $R_{\rm FC}$ as the ratio of the fluorescence over Cherenkov light densities. This quantity is represented as a function of $x$ for the corresponding studied cases in the lower panels of figures \ref{fig:profiles_wide_angle} and \ref{fig:profiles_IACT}. The parameter $R_{\rm FC}$ is very small ($<1\%$) inside the Cherenkov light pool, while it grows roughly exponentially at core distances beyond some hundreds of meters. At around 1000~m, it will be very significant, i.e., $\sim 5$\% for an IACT and even larger than 10\% for a WACD. A detailed quantification of $R_{\rm FC}$ as a function of the observation conditions is given in the next subsections.
\par
Note that our results for IACTs only account for the integrated signal, ignoring its time evolution and how it is distributed over the camera pixels. All the fluorescence and Cherenkov photons detected by an IACT reach the same region of the camera, that is, that delimited by the angular extension of the shower inside the telescope FoV. However, the light is distributed in this region in a different way for both components due to their distinct angular distributions. For an on-axis observation, Cherenkov light is concentrated at small angular distances with respect to the center of the camera (i.e., the upper part of the shower), while fluorescence light is much more uniform in both angular distance and time. A detailed determination of the fluorescence contamination in a particular IACT will require an end-to-end simulation including the detector as well as the analysis of the camera images, which is beyond the scope of this paper. Nevertheless, we made some tests imposing the arrival photon times to be within a time window of 100~ns, which covers the entire range of a typical Cherenkov signal when integrating over all the significant pixels. In doing so, the calculated $R_{\rm FC}$ value at $x=1000$~m for a FoV of $10^\circ$ is slightly reduced from 6\% to 5\%.
        
\subsection{Energy dependence}
\label{subsec:energy}
To the first order, the total number of charged particles in an EAS as well as the energy deposited by them scale
with the energy of the primary particle. Therefore, the total amount of Cherenkov and fluorescence light scales with energy too. However, the spatial distributions of both light components on ground also depend on the longitudinal development of the shower. The higher the shower energy, the deeper into the atmosphere it develops and the more steeply these light distributions decrease with core distance. This effect is stronger in the fluorescence component than in the Cherenkov one, since the latter is concentrated within the pool. As a consequence, the fluorescence contamination should in principle increase with increasing energy. On the other hand, a finite FoV has an opposite effect because it reduces the fluorescence contamination, as explained before. This reduction is more important as the energy increases, because the deeper into the atmosphere the shower develops, the larger the fraction of the shower development outside the telescope FoV.
\par
\begin{figure}[ht]
\begin{center}
\centering\includegraphics[width=0.864\linewidth]{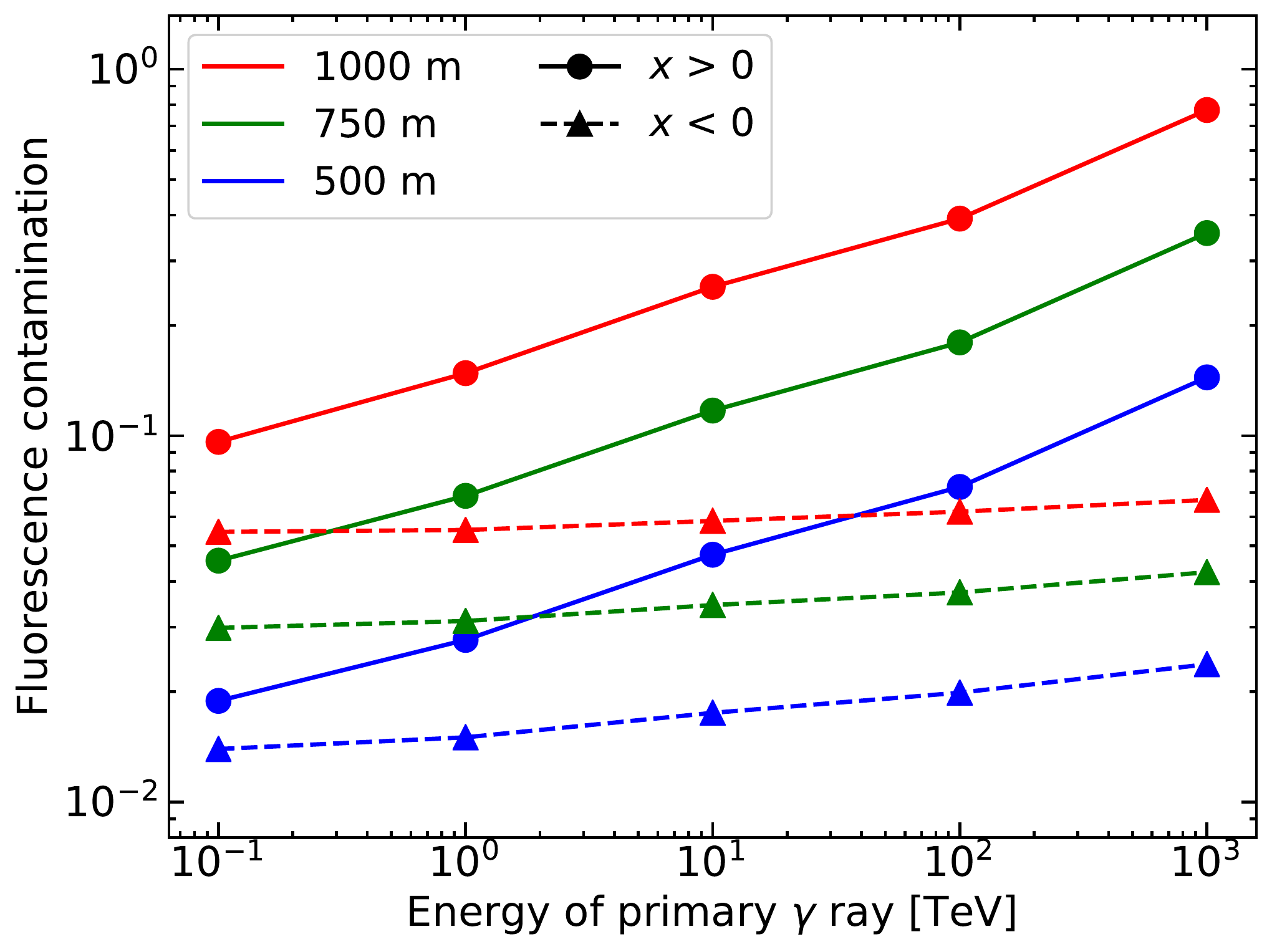}
\caption{Fluorescence contamination $R_{\rm FC}$, as a function of the shower energy and $x$ distance for WACDs with FoV of $60^\circ$ in diameter in the vertical direction. Results were obtained for $\gamma$-ray showers with a zenith angle of $20^\circ$. The statistical error bars are smaller than the symbol size.} 
\label{fig:E_dependence_wide_angle}
\end{center}
\end{figure}

The dependence of the fluorescence contamination on the primary $\gamma$-ray energy for wide-angle detectors ($60^\circ$ FoV in the vertical direction) is shown in figure \ref{fig:E_dependence_wide_angle} for showers with $\theta=20^\circ$ and at the characteristic core distances of 500, 750 and 1000~m. $R_{\rm FC}$ is found to grow approximately as $E^{0.22}$ for positive $x$ values (solid lines), reaching up to $R_{\rm FC}\approx 1$ at 1000~m for 1~PeV showers. The contamination is less significant and weaker dependent on energy for negative $x$ values (dashed lines), because the impact of the FoV is more important in this region, as shown in figure \ref{fig:profiles_wide_angle}, and compensates the above-mentioned effect of the deeper shower development.

\begin{figure}
\begin{center}
\centering\includegraphics[width=0.864\linewidth]{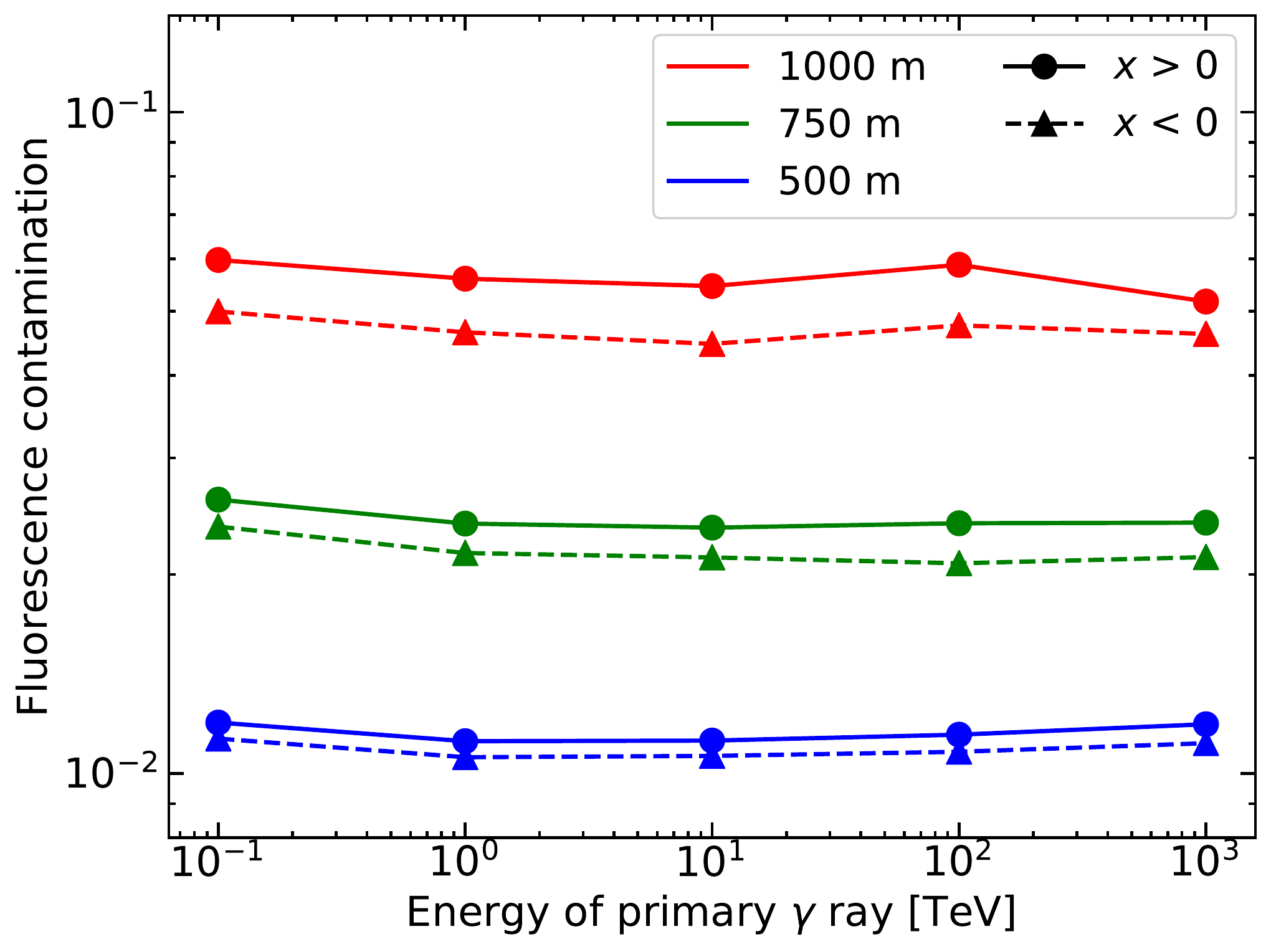}
\caption{The same as figure \ref{fig:E_dependence_wide_angle}, but for IACTs with a FoV of $10^\circ$ in diameter centered at the arrival shower direction.}
\label{fig:E_dependence_IACT}
\end{center}
\end{figure}

Figure \ref{fig:E_dependence_IACT} shows the results for IACTs with a FoV of $10^\circ$, observing $20^\circ$
inclined showers on axis. For this FoV, $R_{\rm FC}$ becomes non-negligible ($>1\%$) at a distance larger than 500~m and is found to be nearly independent of energy at all core distances (the asymmetry between positive and negative $x$ values is very small). Although we report results for a wide range of shower
energies, it should be stressed that, outside the Cherenkov light pool, the photon density for showers with energy lower than 1~TeV is comparable to typical night sky background values, thus the fluorescence contamination is only relevant for higher-energy showers when observed from large core distance.

\subsection{Dependence on shower zenith angle}
\label{subsec:zenith_angle}
\begin{figure}[!htb]
\begin{center}
\centering\includegraphics[width=0.864\linewidth]{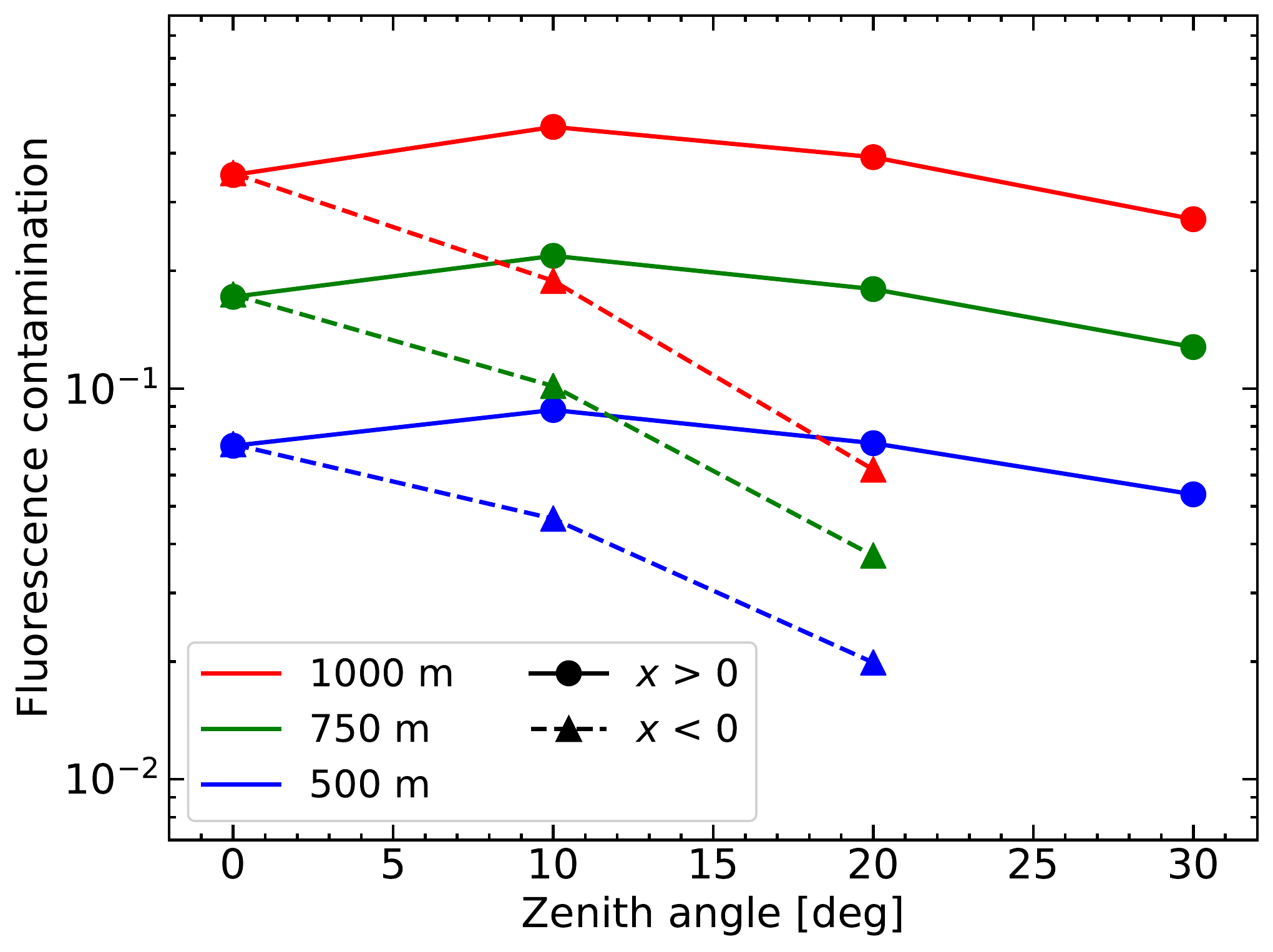}
\caption{Fluorescence contamination as a function of zenith angle and $x$ distance for WACDS with FoV of $60^\circ$ in diameter in the vertical direction. Results were obtained for 100~TeV $\gamma$-ray showers.}
\label{fig:Theta_dependence_wide_angle}
\end{center}
\end{figure}

\begin{figure}[!htb]
\begin{center}
\centering\includegraphics[width=0.864\linewidth]{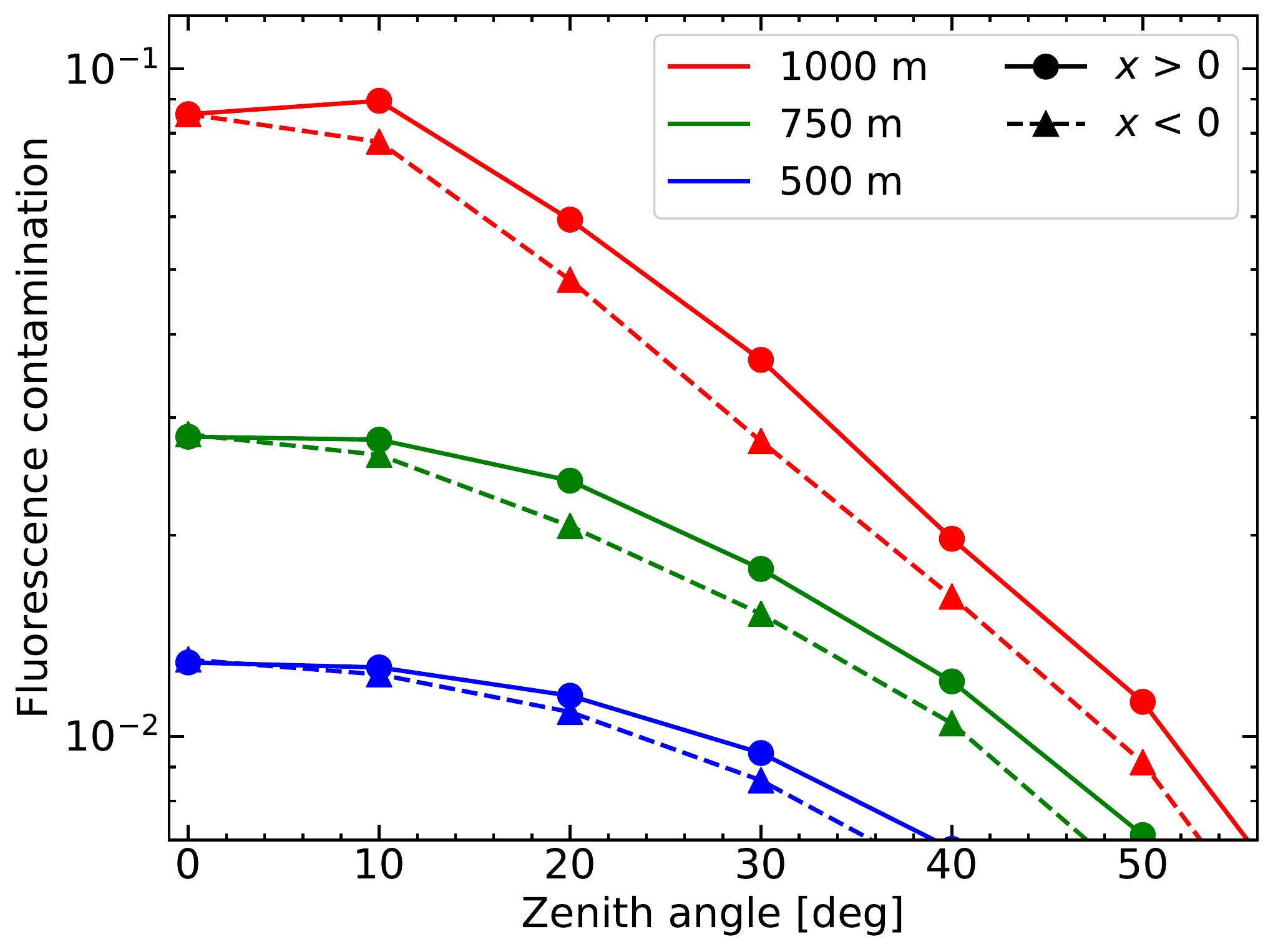}
\caption{The same as figure \ref{fig:Theta_dependence_wide_angle}, but for IACTs with a FoV of $10^\circ$ in diameter centered at the arrival shower direction.}
\label{fig:Theta_dependence_IACT}
\end{center}
\end{figure}

The angular dependence of the fluorescence contamination for 100~TeV $\gamma$-ray showers is shown in figures \ref{fig:Theta_dependence_wide_angle} and \ref{fig:Theta_dependence_IACT} for the two techniques studied in this work. The behavior shown in these figures can be understood in terms of two effects of the shower inclination on the spatial light distributions. In the first place, the size of the Cherenkov light pool on ground is nearly proportional to $\sec^2 \theta$ along the $x$ axis and to $\sec\theta$ along the $y$ axis, that is, in the coordinate system $x'\equiv x\,\cos^2\theta$, $y'\equiv y\,\cos\theta$, it is delimited by a circle of radius $r'\approx 120$~m at 2200~m a.s.l.\footnote{This can be justified assuming that the Cherenkov light beam can be approximated by an inclined cone with the vertex at a fixed height regardless of the angle of inclination.},
where $r'^2=x'^2+y'^2$.
Therefore, $R_{\rm FC}$ will be modulated by the modified coordinates $(x',\,y')$ rather than by $(x,\,y)$. In the second place, an inclined shower develops higher in the atmosphere because, for a given atmospheric thickness transversed, the corresponding vertical depth is smaller by a factor $\cos\theta$. As a consequence of both effects, $R_{\rm FC}$ generally decreases with $\theta$ for a given $x$ value.

For WACDs and positive $x$ values, this decrease in $R_{\rm FC}$ is partly compensated by the fact that the detector observes the shower development at lower altitudes and with larger angle as $\theta$ increases (see top panel of figure \ref{fig:geometry}) \footnote{Note that for $\theta>30^\circ$ and positive $x$ values, the detector would operate in {\it fluorescence mode}, as it would observe the shower transversely.}. The opposite happens for negative $x$ values, resulting in a steeper decrease of the fluorescence contamination with $\theta$. At a given modified radial distance $r'$, $R_{\rm FC}$ will be in between the results obtained for these two extreme cases; namely at $(x'=r',\,y'=0)$ and $(x'=-r',\,y'=0)$. Nonetheless a precise determination of the fluorescence contamination in an actual experiment would require to simulate also the detector.
\par
In the IACT case, the asymmetry between positive and negative $x$ values is very small and the narrow FoV compensates the effects due to variations in the shower development, as explained in the previous subsection. In fact, for a fixed modified radial distance $r'$, $R_{\rm FC}$ is found to be nearly independent of the shower energy as well as of both the zenith and azimuthal angles. For example, for an IACT with a FoV of $10^\circ$ pointing to a certain $\gamma$-ray source, $R_{\rm FC}$ will be around 3\% for any incoming shower that impacts at $r'=750$~m. Note however that, for a very large zenith angle of $60^\circ$, this would involve a core distance of 3000~m along the $x$ axis or 1500~m along the $y$ axis, which is unreachable for the IACT technique. As a simple recipe, a good estimation of $R_{\rm FC}$ at an arbitrary position ($x$, $y$), $\gamma$-ray energy and zenith angle $\theta$ can be obtained from the result given in figure \ref{fig:profiles_IACT} for $20^\circ$ inclined showers at an equivalent core distance
\begin{equation}
x_{\rm eq}\equiv r'\sec^2 20^\circ=1.13\sqrt{\left(x\,\cos^2 \theta \right)^2+\left(y\,\cos \theta \right)^2}\,.
\end{equation}

\subsection{Fluorescence contamination in off-axis observations}
\label{subsec:FoV}
\begin{figure}[!htb]
\begin{center}
\centering\includegraphics[width=0.5\linewidth]{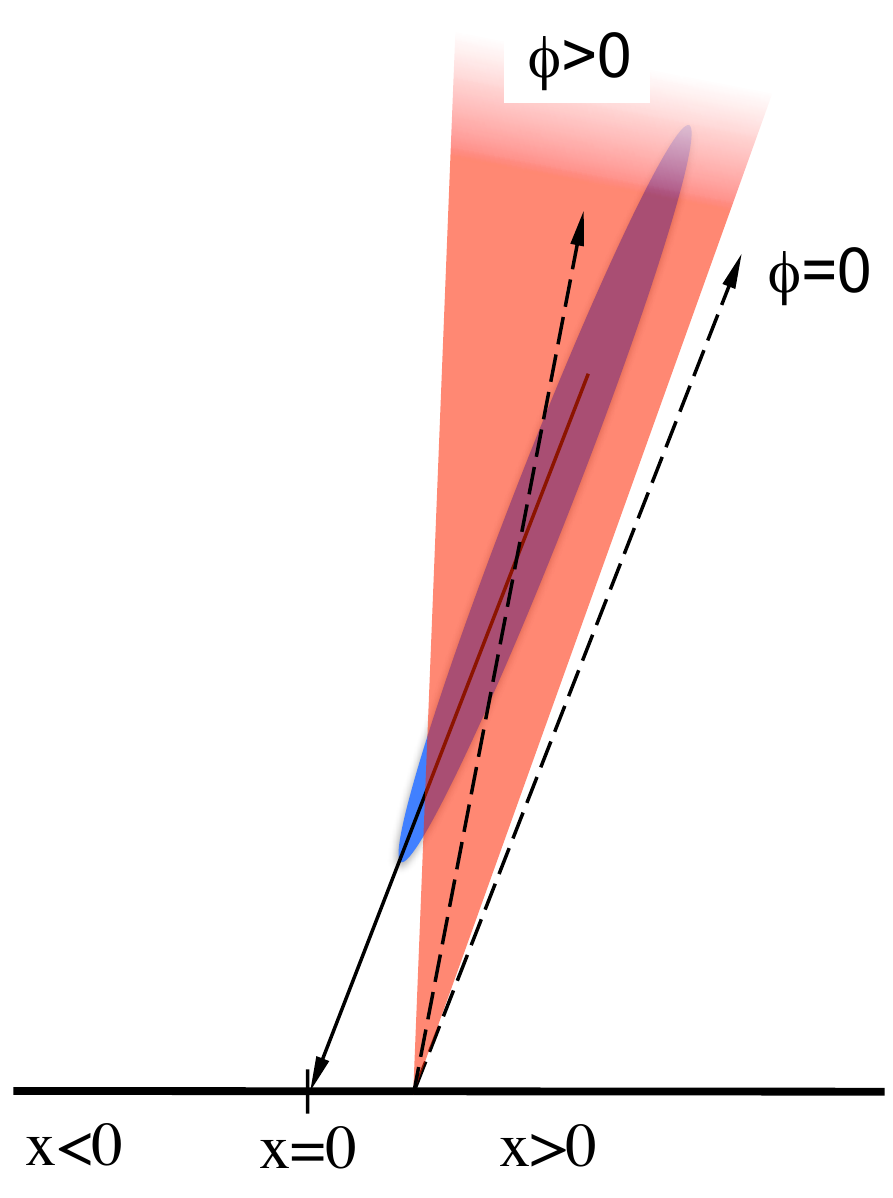}
\caption{Schematic view of an off-axis observation.}
\label{fig:offaxis}
\end{center}
\end{figure}

\begin{figure}[!htb]
\begin{center}
\centering\includegraphics[width=0.864\linewidth]{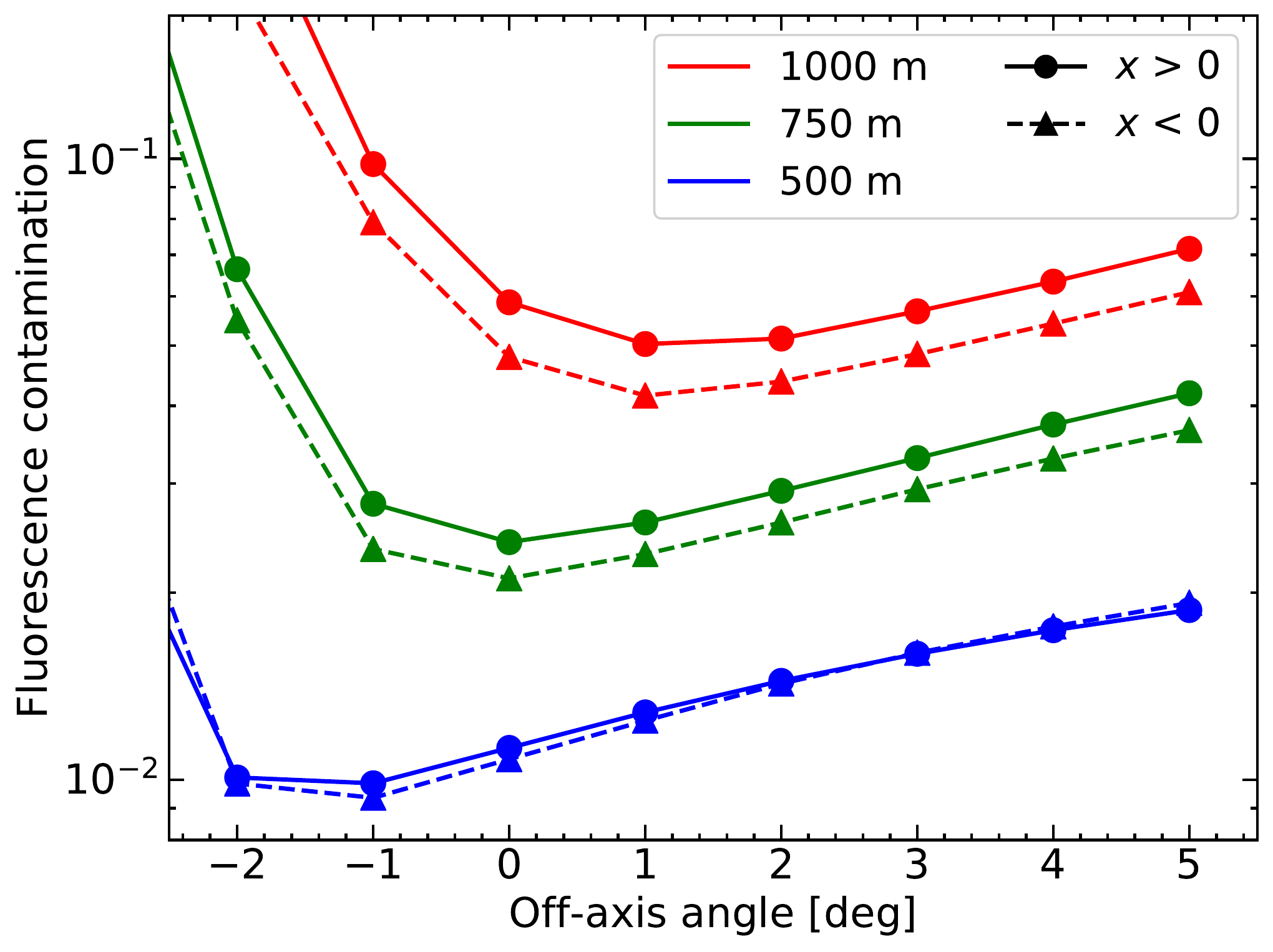}
\caption{Fluorescence contamination as a function of telescope off-axis angle and $x$ distance for IACTs with a FoV of $10^\circ$ in diameter. Results were obtained for 100~TeV showers with $20^\circ$ zenith angle.}
\label{fig:offset_dependence}
\end{center}
\end{figure}

Quite often IACTs do not point exactly in the direction of the $\gamma$-ray source, but observe the incoming showers with an off-axis angle $\phi$ of up to a few degrees, as illustrated in figure \ref{fig:offaxis}. This is the case of sky surveys (e.g., \cite{Abdalla2018}), observation of extended sources, etc. We show here that the fluorescence contamination may be larger for off-axis observations.
As a representative case, the lateral light profiles for the sample of 100~TeV showers with $\theta=20^\circ$ and FoV of $10^\circ$ (figure \ref{fig:profiles_IACT}) have been recalculated for several off-axis angles $\phi$. The results on the fluorescence contamination are presented in figure \ref{fig:offset_dependence}. For simplicity, we assumed that both the shower axis and the main optic axis of the telescope are on the $XZ$ plane ($\phi$ is the angle between them). Positive $\phi$ values indicate the case in which the two axes intersect at a point above ground level, i.e., the case illustrated in figure \ref{fig:offaxis}. This is equivalent to augmenting the telescope FoV in an on-axis observation, because it increases the observed fraction of the shower development.\footnote{Note however that further increasing $\phi$ makes the telescope to miss the upper part of the shower development. An extreme value of $\phi=90^\circ$ would correspond to observing the shower transversely.} Therefore, the $R_{\rm FC}$ value increases with $\phi$. A negative $\phi$ value corresponds to the opposite situation in which the IACT only samples the upper part of the shower. For large core distances of interest for this work, this involves that the amount of collected Cherenkov light may be insufficient to record a useful image.

In the general case that the telescope axis is outside the $XZ$ plane, the fluorescence contamination for small $\phi$ angles is expected to be moderately larger with respect to the on-axis observation. It should be noted, however, that IACTs with a FoV larger than $10^\circ$ (e.g., LHAASO-WFCTA \cite{Zhang2011-LHAASO}), may have a very relevant fluorescence contamination in case they can observe with $\phi>5^\circ$.

\subsection{Cross-check with a numerical algorithm}
\label{subsec:algorithm}

Our implementation of the fluorescence production in CORSIKA has been tested against a numerical algorithm \cite{ICRC2015,ICRC17} developed for that purpose. This algorithm does not include statistical fluctuations in the photon production and assumes a one-dimension shower development, in a similar way to other codes used for the evaluation of the fluorescence emission in UHECR showers (e.g., \cite{PradoJr.2005}).
\par
The cross-check between the simulation results and these numerical calculations has been carried out with vertical $\gamma$-ray showers of 100~TeV and assuming a FoV of $10^\circ$ around the vertical direction. Simulation results were averaged over 100 showers. The longitudinal profile of energy deposit used in the numerical algorithm was the averaged one from the same MC sample.
\par
\begin{figure}[!htb]
\begin{center}
\centering\includegraphics[width=0.864\linewidth]{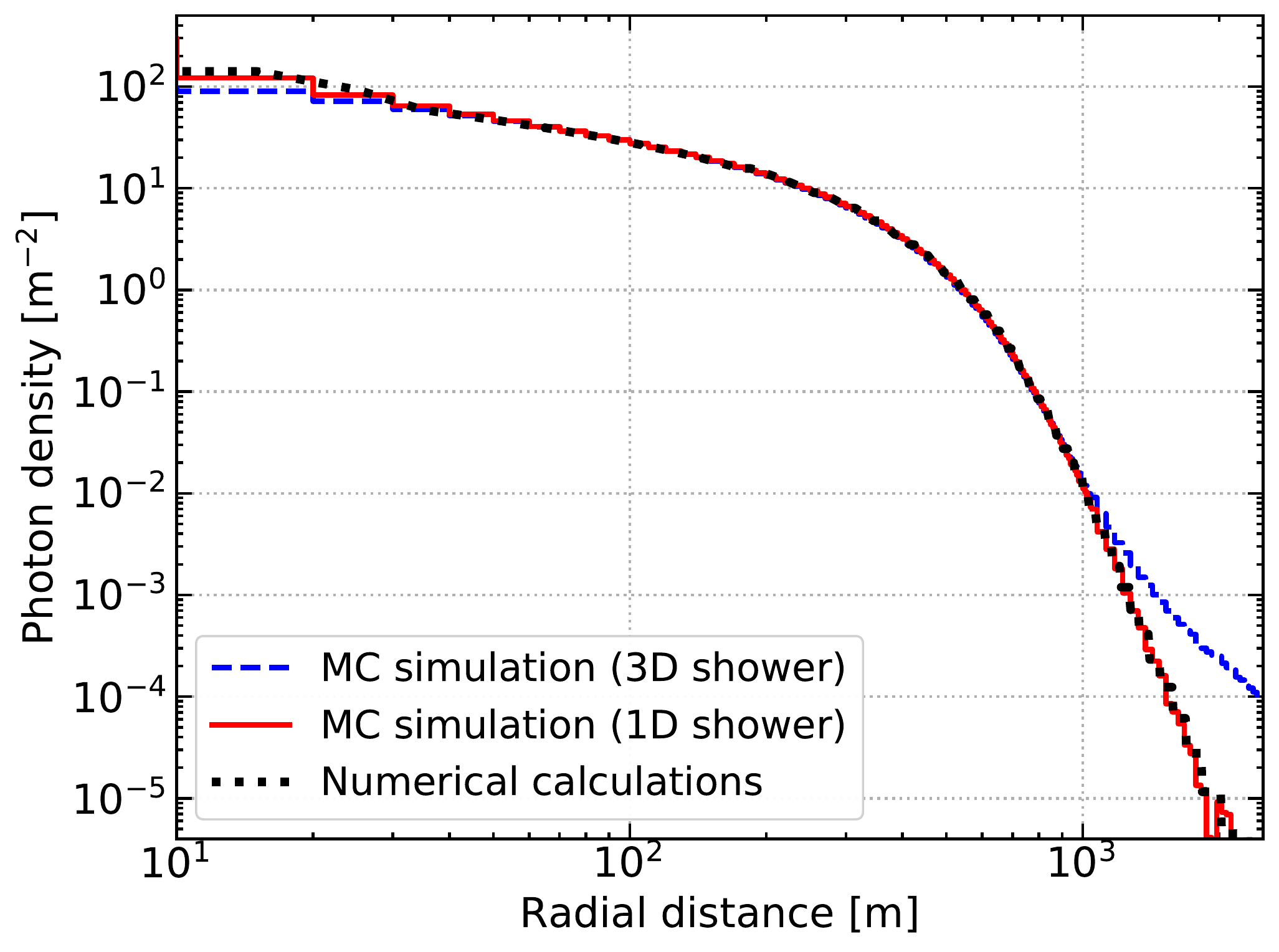}
\caption{Comparison between the simulated fluorescence photon radial distribution from 100~TeV $\gamma$-ray vertical showers and the one obtained from numerical calculations. An excellent agreement is found when forcing a one-dimension shower development in CORSIKA.}
\label{fig:comparison}
\end{center}
\end{figure}

The comparison of radial distributions of fluorescence photons evaluated by both procedures is shown in figure \ref{fig:comparison}. As can be seen, the numerical calculations (dotted line) overestimate the fluorescence photon density at very small core distance, while underestimate it at very large $r$ values. These discrepancies can be attributed to the linear-shower approximation applied in our algorithm. To prove this, such approximation was replicated in CORSIKA by modifying the output so that the arrival bunch positions on ground were calculated as if they were generated at the shower axis. The results are represented by solid lines in the plot, showing an excellent agreement with those of the numerical algorithm.

\section{Future work}
\label{sec:future_work}

We aim to include the implementation of fluorescence emission in the official version of CORSIKA software. To that end, there are still some upgrades to be done as explained at the end of section \ref{sec:CORSIKA}. This tool is operative for both electromagnetic and hadronic showers. Thus, it would allow $\gamma$-ray and cosmic-ray communities to simulate both the Cherenkov and fluorescence light contributions to the recorded signals taking into account all the particularities of their instruments (e.g., the detection efficiency, the ray-tracing in the telescope, the time evolution of signals and the setting trigger conditions). Also it will be possible to account for the atmospheric conditions. This may be specially interesting when both light components are comparable.
\par
In fact, we plan to extend this work in the future for the IACT technique, quantifying in more detail the impact of the fluorescence contamination in the camera images as well as in the reconstruction of both the energy and the arrival direction of showers. This will require the use of dedicated Monte Carlo simulation tools for IACTs (e.g., sim\_telarray \cite{Bernlohr2008}) and associate analysis software.
\par
The numerical algorithm used to validate our implementation of fluorescence in CORSIKA can also evaluate the Cherenkov light production from EASs. All the details of this complementary tool, previously outlined in \cite{ICRC2015,ICRC17}, will be published soon. It provides quick but reasonable predictions on the detected light intensity at arbitrary conditions and shower energies, even including some effects that may be costly to simulate. In particular, we used it to estimate the fluorescence contamination when including the wavelength-dependent atmospheric absorption and the detection efficiency of a typical Cherenkov telescope. We found that the $R_{\rm FC}$ values given in section \ref{sec:results} may be reduced by a factor of $\sim 0.8$ after correcting for these effects. 
\par
In a previous work \cite{ICRC2015}, we used the above numerical algorithm to give, for the first time, preliminary results on the capability of a large array of IACTs, like those of the CTA Observatory, to detect EASs using the fluorescence technique. This operation mode would benefit from stereoscopic observations with a potentially large effective area, and it could be used simultaneously to regular observations with the Cherenkov technique. A single IACT with a typical FoV of $10^\circ$ can track about 5~km of the longitudinal development of a shower when observed transversely from a distance greater than 30~km. This might allow the detection of UHE cosmic rays as long as signals can be recorded within a time window of tens of microseconds. For smaller shower-to-telescope distances (hence, smaller shower energies), the fraction of shower development observed would be insufficient for a full reconstruction. Nevertheless, this technique could still be used if several telescopes track different parts of the shower development, with the advantage that, in contrast to the fluorescence telescopes, their higher angular resolution may provide additional information on the shower lateral development. We are upgrading our numerical algorithm to include realistic trigger conditions of telescopes taking into account the lateral shower development, among other things. This work will be published in a paper in preparation.

\section{Conclusions}
\label{sec:conclusions}

We have implemented the atmospheric fluorescence emission in the CORSIKA code in a similar way to the Cherenkov light production. Updated air fluorescence data and models were used for this purpose. With the assistance of this tool we have quantified the fluorescence contribution to the signals recorded by ground-based air-Cherenkov telescopes, which is so far ignored in the reconstruction of EASs. 
\par
As a consequence of the very different angular distributions of these two light components, the Cherenkov one dominates within the light pool. In addition, the fluorescence contribution is strongly suppressed in telescopes with a narrow FoV, like IACTs. However, the fluorescence contamination is $\sim 1$\% at a core distance of 500~m and significant ($>5$\%) at 1000~m. 
Thus ignoring this effect would have an impact on VHE $\gamma$-ray observations carried out by arrays of IACTs that collect measurable signals at large core distances (e.g., CTA). Note that the fluorescence contamination is significantly larger than the one from the Cherenkov radiation scattered in the atmosphere \cite{Bernlohr2000}, which is usually neglected.
\par
It has been found that the fluorescence contamination in IACTs is nearly independent of the shower energy and weakly dependent on the zenith angle for a fixed distance relative to the Cherenkov light pool. As a consequence, for very inclined showers ($\theta > 50^\circ$), this is negligible at a core distance smaller than 1000~m. We also found that off-axis observations lead to an increase in the fluorescence contamination. 
\par
According to our simulations, arrays of WACDs are strongly affected by the fluorescence contamination, in particular, for the highest energy showers. In the PeV region, the signals of counters located 1000~m from the shower core can contain up to 45\% of fluorescence (i.e., $R_{\rm FC} = 0.8$).
\par
As a future work, we plan to include our implementation of fluorescence emission in the official version of the CORSIKA code, making it available for the detailed simulation of IACTs and arrays of WACDs working at conditions where fluorescence contamination is expected to be non-negligible. In addition, this tool will allow us to pursue a detailed study on the feasibility of using an array of IACTs in fluorescence mode.

\section*{Acknowledgments}
We gratefully acknowledge support from Spanish MINECO (contracts FPA2015-69210-C6-3-R and FPA2017-82729-C6-3-R) and the European Commission (E.U. Grant Agreement 653477). D. Morcuende acknowledges a predoctoral grant UCM-Harvard University (CT17/17--CT18/17) from \textit{Universidad Complutense de Madrid}.

\bibliography{fluor.bib}

\end{document}